\documentclass[showpacs,aps,prl,twocolumn,epsfig,superscriptaddress]{revtex4}
\usepackage{amssymb}

\usepackage{graphicx}

\begin{document}

\title{Critical temperature and giant isotope effect in presence of
paramagnons.}

\author{ O.V. Dolgov}

\affiliation{Max-Planck-Institut f\"{u}r Festk\"{o}rperphysik, Heisenbergstr.1,
70569 Stuttgart, Germany}

\author{I.I. Mazin }
\affiliation{Center for Computational Materials Science, Naval
Research Laboratory, Washington, DC 20375, USA}

\author{A.A. Golubov}
\affiliation{Faculty of Science and Technology, University of
Twente, 7500 AE Enschede, The Netherlands}

\author{S.Y. Savrasov}
\affiliation{Department of Physics, New Jersey Institute of
Technology, Newark, New Jersey 07102, USA}

\author{E.G. Maksimov}
\affiliation{P.N. Lebedev Physical Institute, RAS, Leninskii pr. 53, 119991
Moscow, Russia}

\begin{abstract}
We reconsider the long-standing problem of the effect of spin fluctuations
on the critical temperature and isotope effect in a
phonon-mediated superconductor. Although the general physics of the
interplay between phonons and paramagnons had been rather well understood,
the existing approximate formulas fail to describe the correct behavior of $%
T_{c}$ for general phonon and paramagnon spectra. Using a
controllable approximation, we derive an analytical formula for
$T_{c}$ which agrees well with exact numerical solutions of the
Eliashberg equations for a broad range of parameters. Based on both
numerical and analytical results, we predict a strong enhancement of
the isotope effect when the frequencies of spin fluctuation and
phonons are of the same order. This effect may have important
consequences for  near-magnetic superconductors such as MgCNi$_{3}$.
\end{abstract}

\date{\today }
 \maketitle

In the last decade a large number of superconductors were discovered
in which enhanced spin fluctuations (SF) play a role in the
superconductivity, $e.g.,$ Sr$ _{2}$RuO$_{4},$ MgCNi$_{3},$
$\varepsilon $-Fe, ZrZn$_{2},$ and others, bringing about new and
interesting physics. However, understanding such materials, even at
an intuitive level, has been hindered by the lack of a simple
formula that would approximate the full Eliashberg theory in a
compact analytical form, as the conventional McMillan formula (MMF)
does. As a result, uncritical generalizations of the latter have
been used as a substitute, despite the fact that, as we will show
below, some of them are too approximate or outright incorrect. In
this Letter we present an analogue of the MMF, derived in a
controllable way and tested against numerical solutions of full
Eliashberg equations, including interaction with SF (paramagnons).
We point out the possibility of a giant \textit{phonon }isotope
effect induced \textit{by SF}. We will also apply this theory, as an
example, to a nearly-ferromagnetic superconductor, MgCNi$_{3}.$

The understanding that SF are pair breakers in conventional
superconductors is nearly as old as the BCS theory itself
\cite{BSGJS}. Moreover, it was soon realized that strong coupling
manifests itself in a nontrivial way in the presence of SF
\cite{Rietschel,VIS}. In a number of papers numerical solutions of
the Eliashberg equations were presented, incorporating phonon
$\alpha ^{2}F_{p}(\omega )$ as well as SF $\alpha ^{2}F_{s}(\omega
)$ spectral functions (see, e.g., Refs. \cite{WCNC}). However,
solving the full Eliashberg equation is not always an option, and
does not provide as much physical insight as analytical treatment.
An analytical tool comparable to the famed MMF is needed.

Retrospectively, one can realize that the overwhelming success of
the MMF is due to three facts: (a) it can be derived analytically
using simple approximations, (b) it includes Coulomb repulsion
effects, (c) it has three universal adjustable parameters, which,
after little tuning, produce an expression which is surprisingly
accurate for a large range of phonon frequencies and coupling
strengths. Compared to the BCS equation, the MMF includes three
essential pieces of additional physics: effective mass
renormalization, logarithmic reduction of the Coulomb repulsion, and
proper (logarithmic) averaging of the phonon frequency. All three
effects can be derived analytically in some approximations. In fact,
it is known that the functional form of the MMF can be derived in
two different ways. One, known as the square-well model
\cite{AllenMitr}, uses
 the Matsubara representation, where the coupling with the
phonons is parametrized in terms of the matrix $\lambda (n,n^{\prime
}).$ The model assumes two different approximations for the same
function $ \lambda (n,n^{\prime }),$ depending on whether it is used
in the equation for the mass renormalization $Z$ or in the one for
the gap function $\phi$:

\begin{eqnarray}
\lambda _{Z}(n,n^{\prime }) &=&\lambda _{p}\Theta (\omega _{p}-\left| \omega
_{n-n^{\prime }}\right| )\  \\
\lambda _{\phi }(n,n^{\prime }) &=&\lambda _{p}\Theta (\omega _{p}-\left|
\omega _{n}\right| )\Theta (\omega _{p}-\left| \omega _{n^{\prime }}\right|
).  \nonumber
\end{eqnarray}%
This models leads to an equation for the critical temperature,
$T_{c},$
\begin{equation}
T_{c}=a\omega _{\log }\exp \{-b(1+\lambda _{Z})/[\lambda _{\phi }-\mu ^{\ast
}(1+c\lambda _{Z})],  \label{MM0}
\end{equation}%
where the theoretical parameters are $a=1.14,$ $b=c=1$, $\lambda
_{Z}=\lambda _{\phi }=\lambda _{p}=2\int_{0}^{\infty }\omega
^{-1}\alpha ^{2}F_{p}(\omega )d\omega $ and $\lambda _{p}\ln \omega
_{\log }=2\int_{0}^{\infty }\omega ^{-1}\ln \omega \alpha
^{2}F_{p}(\omega )d\omega $. The renormalized Coulomb potential is
reduced from its bare value $\mu $ as $\mu ^{\ast }=\mu /(1+\mu \ln
\frac{\omega _{C}}{\omega _{\log }}),$ where $\omega _{C}$
characterizes the frequency cutoff of the Coulomb interaction. The
MMF formula is given by Eq.\ref{MM0} with optimized parameters
$a=1/1.2,$ $b=1.04,$ and $c=0.62.$

SF, as opposed to phonons, induce repulsion for singlet pairs.
However, they contribute to the mass renormalization just the same.
Therefore the first instinct is to let $\lambda _{Z}=\lambda
_{p}+\lambda _{s},$ where $\lambda _{s}$ describes the SF, and
$\lambda _{\phi }=\lambda _{p}-\lambda _{s}.$ Eq.\ref{MM0} with this
modification and standard $a,b$ and $c$ is the one routinely used in
the literature for materials with SF (e.g.,
Refs.\cite{MazinFe,helge,someJarlborg}).

Obviously, using two different approximations for the same physical
function $\lambda (n,n^{\prime })$ depending on whether it appears
in the first or second Eliashberg equation cannot be justified by
any logic. It \textit{appears} that the MMF formula can be \textit{%
fortuitously} derived in this way, but, as we will see below, this
approach fails when SF are included. An alternative derivation of
the MMF utilizes the real frequency axis formalism
\cite{kmmProblema}. The one-mode approximation is
used, which assumes an Einstein phonon at a frequency $\omega _{p},$ \textit{%
i.e., }$\alpha ^{2}F(\omega )=\lambda _{p}\omega _{p}\delta (\omega
-\omega _{p})/2.$ The Eliashberg equations are then solved
iteratively. After the first iteration one obtains
\cite{kmmProblema}
\begin{equation}
T_{c}=1.14\omega _{p}\exp \left\{ -\frac{1}{2}-\frac{1+\lambda _{p}}{\lambda
_{p}-\mu ^{\ast }[1+0.5\frac{\lambda _{p}}{1+\lambda _{p}}]}\right\} ,
\end{equation}%
which is similar to the square well formula Eq.\ref{MM0} with $a=1.14/\sqrt{e%
}=1/1.44$ (note that this value of $a$ is much closer to the optimized one)$,
$ $b=1,$ and $c=0.5/(1+\lambda _{p})$. This approach is a controllable
approximation with a concrete physical meaning. However, it has never been
applied to superconductors with SF.

On the contrary, several attempts to apply the square well model to SF have
been reported. In Refs. \cite{VIS,WCNC} the following expression was derived
(for $\mu ^{\ast }=0$):
\begin{eqnarray}
T_{c}&=&1.14\omega _{p}^{\nu }\omega _{s}^{1-\nu }\exp \{-(1+\lambda
_{p}+\lambda _{s})/(\lambda _{p}-\lambda _{s})], \label{MM2}  \\
{\rm with } &\nu& =\lambda _{p}/(\lambda
_{p}-\lambda _{s}) \ \ \ \ \ \ \ \ \ \ \ \ \ \ \ \ \ \ \ \ \ \ \ \ \ \ \ \
\ \ \ \ \ \ \ \ (a) \nonumber \\
{\rm or} &\nu& =%
\frac{\lambda _{p}^{2}}{\lambda _{p}-\lambda _{s}}\left[ \lambda
_{p}-\lambda _{s}+\frac{\lambda _{p}\lambda _{s}}{1+\lambda _{p}+\lambda _{s}%
}\ln \frac{\omega _{p}}{\omega _{s}}\right]^{-1} \ (b)  \nonumber
\end{eqnarray}%
where the choice (a) is due to Carbotte \textit{et al}. \cite{WCNC},
and (b) to Vonsovskii \textit{et al}. \cite{VIS}. Unfortunately,
neither authors give details of their derivations, so we do not know
what was different in their models. We were not able to reproduce
either result. The latest paper utililizing the square well model
(in the weak coupling limit) is that by Shimahara \cite{Shim}. Our
own result for the square well model reduces to that of
Ref.\cite{Shim} in the weak limit, and reads
\begin{eqnarray}
T_{c} &=&1.14\omega _{p}\exp [-\frac{1+\lambda _{s}+\lambda _{p}}{\lambda
_{p}-\frac{\lambda _{s}(1+\lambda _{s})}{1+\lambda _{s}+\lambda _{s}\ln
\frac{\omega _{s}}{\omega _{p}}}}]~\mathrm{,~}\omega _{s}\geq \omega _{p}
\label{tcA} \\
T_{c} &=&1.14\omega _{s}\exp [-\frac{1+\lambda _{s}+\lambda _{p}}{\frac{%
\lambda _{p}(1+\lambda _{p})}{1+\lambda _{p}-\lambda _{p}\ln
\frac{\omega _{p}}{\omega _{s}}}-\lambda _{s}}]~\mathrm{,~}\omega
_{s}\leq \omega _{p} \label{tcB}
\end{eqnarray}%
Unlike Eq.\ref{MM2}, Eqs.\ref{tcA}, \ref{tcB} reduce to the McMillan form
upon substitution $\omega _{s}\rightarrow \omega _{C}\gg \omega _{p},$ $%
\lambda _{s}\rightarrow \mu ,$ as it should.

Given the controversy about the square-well model, it is desirable to have a
derivation in a controllable approximation, such as the real frequency axis
formalism of Refs.\cite{kmmProblema}. Assuming an Einstein phonon at a
frequency $\omega _{p}$ and an ``Einstein''\ paramagnon at $\omega _{s}$, $%
2\alpha ^{2}F(\omega )=\lambda _{p}\omega _{p}\delta (\omega -\omega
_{p})-\lambda _{s}\omega _{s}\delta (\omega -\omega _{s}),$ we obtain the
following iterative solution of the Eliashberg equations. 
\begin{eqnarray}
T_{c} &=&1.14\omega _{p}^{\frac{\lambda _{p}}{\lambda _{p}-\lambda _{s}}%
}\omega _{s}^{-\frac{\lambda _{s}}{\lambda _{p}-\lambda _{s}}}\exp (K)
\nonumber \\
&\times &\exp \left\{ -\frac{1+\lambda _{p}+\lambda _{s}}{\lambda
_{p}-\lambda _{s}-\mu ^{\ast }(1-K\frac{\lambda _{p}-\lambda _{s}}{1+\lambda
_{p}+\lambda _{s}})}\right\}   \label{Tc} \\
K &=&-\frac{1}{2}-\frac{\lambda _{p}\lambda _{s}}{\left( \lambda
_{p}-\lambda _{s}\right) ^{2}}\left[ 1+\frac{\omega _{p}^{2}+\omega _{s}^{2}%
}{\omega _{p}^{2}-\omega _{s}^{2}}\ln \frac{\omega _{s}}{\omega _{p}}\right]
.  \nonumber
\end{eqnarray}%
For $\omega _{p}\rightarrow \omega _{s}$, $K=-1/2,$ and at $\mu
^{\ast }=0$, Eq.\ref{Tc} reduces to Eq.\ref{MM2} with $\nu =\lambda
_{p}/(\lambda _{p}-\lambda _{s})$.

\begin{figure}[tbp]
\includegraphics[width=2.8in ]{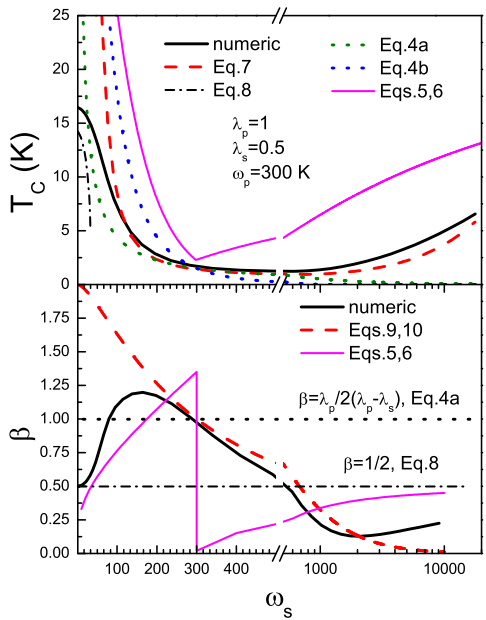}
\caption{Comparison of $T_c$ and isotope coefficient with the exact
numerical calculations. (color on line)} \label{Tcfig}
\end{figure}

As usual, the ultimate test for any approximation is numerical
calculations. We solved the Eliashberg equations for a variety of
model $\alpha ^{2}F(\omega )$ including SF and compare them with the
proposed analytical formulas. In Fig.\ref{Tcfig} we show this
comparison for the simplest "one-mode" approximation, one phonon and
one paramagnon (we have verified that other model spectra lead to
similar results). As we can see, while the Eq.\ref{Tc}, as well as
its simplified version Eq.\ref{MM2}(a), describe the numerical
results rather well when $\omega _{s}$ and $\omega _{p}$ are
comparable, the latter fails at $\omega _{s}\gg \omega _{p},$ and
both fail at $ \omega _{s}\ll \omega _{p}.$ Both effects can be
easily understood: Eq.\ref {MM2} includes $\omega _{s}$ in a
negative power in all regimes, thus leading to a total suppression
of superconductivity at $\omega _{s}\rightarrow \infty .$ In
reality, in this limit the negative effect of the SF is renormalized
down logarithmically in the same spirit as the Coulomb repulsion.

Eqs. \ref{Tc}, \ref{MM2} diverge at $\omega _{s}\rightarrow 0$. This is due
to the fact that the derivations above assume that $\omega _{s},\omega
_{p}\gtrsim \pi T_{c}$. It is possible to treat this regime separately. If $%
\omega _{s}\ll T_{c},$ the SF act as static magnetic defects, and
the standard theory of the magnetic pair-breaking\cite{AG} can be
applied. In the Matsubara representation, at $\omega _{s}=0$ one
needs only to keep the term with $n=m$ in $\lambda _{s}(\omega
_{n}-\omega _{m})$. Then the equations reduce to the standard form
\cite{AG,AllenMitr} with the pair-breaking parameter $\gamma \equiv
(1/2\tau _{P})/\pi T_{c}=\lambda _{s}$. In the weak coupling limit,
$T_{c}$ is
\begin{equation}
T_{c}=T_{c0}\exp [\psi (1/2)-\psi (1/2+\gamma )],  \label{AG}
\end{equation}%
where $T_{c0}=T_{c}(\lambda _{s}=0).$
One important difference exists between pair-breaking by SF with $\omega
_{s}=0$ and by magnetic impurities: in the former case
 the pair-breaking parameter $\gamma $ now does not depend
on $T_{c}$. This has consequences for the isotope effect, as we will see
below.

For small, but finite $\omega _{s}\ll \pi T_{c}$ summation of $\lambda
_{s}(n-m)$ over $n-m$ provides the expression for the pair breaking rate in
Eq.\ref{AG}: $\gamma =\lambda _{s}\frac{T_{c}}{2\omega _{s}}\coth \frac{%
T_{c}}{2\omega _{s}}$. This result coincides with Eq. 5.8 of Ref.\cite%
{Millis} for dynamical pair breaking in anisotropic superconductors
if the anisotropy parameter $g$ (as defined in Ref. \cite%
{Millis}) is set to -1. When $\omega _{s}$ increases, $T_{c}$ drops
sharply with a complete loss of superconductivity at $\omega
_{s}=\omega _{s}^{\ast
}=e^{-C}T_{c0}/2\gamma $ (where $e^{C}\simeq 1.78$). However, the condition $%
\omega _{s}\ll \pi T_{c}$ used in the derivation of Eq.\ref{AG} is
lost well before $\omega _{s}^{\ast }$ (in fact, at
$\omega_{s}\simeq \omega _{s}^{\ast }/2)$.

One can take into account the strong coupling effects in the
square-well model, resulting in a renormalization $\gamma
\rightarrow \gamma /(1+\lambda _{p})=\frac{\lambda _{s}}{1+\lambda
_{p}}\frac{T_{c}}{2\omega _{s}}\coth \frac{T_{c}}{2\omega _{s}}.$ As
the comparison with numerical calculations shows (Fig. \ref{Tcfig}),
this approximation underestimates $T_{c}$. However, it illustrates
why $T_{c}$ flattens out at a finite value smaller than $T_{c0}$,
when $\omega _{s}\rightarrow 0$, instead of raising as Eq.\ref{Tc}
suggests.

We also show in Fig.\ref{Tcfig} that both Eq.\ref{MM2}(b) and the square
well model, Eqs.\ref{tcA} and \ref{tcB}, disagree qualitatively with the
numerical results in the whole range of $\omega _{s}$.

We will now turn to the isotope effect. Looking at Eq.\ref{MM2}, one
observes that the isotope coefficient, $\beta =\nu /2=\lambda
_{p}/2(\lambda _{p}-\lambda _{s})>0.5,$ is always enhanced compared
to its BCS value and is independent of the SF frequencies. Clearly,
this should hold approximately in the range of the applicability of
this formula, $\omega _{s}\simeq \omega _{p}\gg \pi T_{c}.$ Indeed,
the more accurate Eq.\ref{Tc} yields for $\beta $
\begin{eqnarray}
\beta  &=&0.5\frac{\lambda _{p}}{\lambda _{p}-\lambda _{s}}\left[ 1-\frac{%
\lambda _{s}}{\lambda _{p}-\lambda _{s}}F\left( \frac{\omega _{s}^{2}}{%
\omega _{p}^{2}}\right) \right]  \\
F(r) &=&({r^{2}-2r\ln r-1})/{(r-1)^{2}} \nonumber.
\end{eqnarray}%
The second term here is the correction to Eq.\ref{MM2}. It can be of
either sign, since with growing $r$ the $F(r)$
 monotonically grows from -1 to
1, and  $F(1)=0.$ As discussed, Eq.\ref{MM2} itself becomes invalid
at $\omega _{s}<\pi T_{c}.$ As $\omega _{s}\rightarrow 0$, according
to Eq.\ref{AG},  $\beta =0.5$ (note that in the case of magnetic
impurities $\beta >0.5$ due to the dependence of $\gamma $ on
$T_{c}$ \cite{AllenMitr}). Therefore, the isotope effect has to have
a maximum at some $0<\omega _{s}<\omega _{p},$ and $\beta _{\max
}>\lambda _{p}/2(\lambda _{p}-\lambda _{s}).$

This is confirmed by numerical calculations, which do show that the
maximum isotope effect for given $\lambda _{s},\lambda _{p}$ is
achieved close to $\omega _{s}\sim \omega _{p}$ and is not far from
$\lambda _{p}/2(\lambda _{p}-\lambda _{s}).$ This is a very import
result, and we emphasize it again: \textit{if superconductivity is
depressed by spin-fluctuations, the total isotope effect increases
compared to its BCS value.}

We shall now apply this formalism to a superconductor where $T_{c}$ is
believed to be substantially suppressed by SF, MgCNi$_{3}$ \cite%
{we,kitaicy,helge}, which has attracted substantial interest not
because of its relatively modest critical temperature, $T_{c}\approx
8$ K, but because of its unusual antiperovskite crystal structure
and proximity to ferromagnetic instability. The latter was first
pointed out by Rosner \textit{et al.} \cite{helge}, who believed in
such strong coupling with SF that they proposed a $p-$ wave
superconductivity. Singh and Mazin \cite{we} also came to the
conclusion that SF should play a role in superconductivity of
MgCNi$_{3},$ but, based on their frozen phonon calculation, they
deduced a large electron-phonon coupling constant ($\lambda
_{p}\gtrsim 1)$ due to the bond-bending Ni phonons. They reconciled
this relatively large $\lambda _{p}$ with a modest $T_{c}$ within a
scenario of s-wave phonon-induced superconductivity depressed by SF.
Later this scenario was re-invented by Shan \textit{et al.}
\cite{kitaicy}, who proved the s-symmetry of the order parameter by
tunneling experiments. This point has been since confirmed by
several groups and seems to be well established.

Singh and Mazin's \cite{we} prediction of the Ni phonon playing the
major role in the electron-phonon coupling in MgNiC$_{3}$ was based
on a limited number of calculations at a high-symmetry point in the
Brillouin zone, and therefore was more an educated guess than a
quantitative argument. A quantitative analysis was provided by
Ignatov \textit{et al.} \cite{serega}, who performed linear-response
calculations of the phonon frequencies and their coupling with
electrons for the whole Brillouin zone. They found a gigantic
coupling for the Ni bond-bending modes, and the most strongly
coupled modes (the mode considered by Singh and Mazin was not among
them) actually unstable. In other words, they found a set of
double-well type instabilities involving mostly Ni atoms. This was
verified by EXAFS measurements \cite {serega}. Ignatov \textit{et
al.} \cite{serega} estimated the total electron-phonon coupling
constant as 1.5 and the logarithmically averaged phonon frequency as
131 K.

Thus, the scenario of Ref. \cite{we} was modified in Ref.
\cite{serega} in the sense that electron-phonon coupling and
superconductivity were coming from highly anharmonic predominantly
Ni modes, but not exactly the simple rotations of the Ni$_{6}$
octahedra considered in Ref. \cite{we}. Unfortunately, strong
anharmonicity of these modes makes it impossible to evaluate their
coupling with electrons in the linear response calculations, but it
is obviously strong. However, one can estimate the electron-phonon
and electron-paramagnon coupling indirectly from experimental data.
Indeed, specific heat renormalization, from different reports,
ranges from 2.6 to 3.1 (see Ref. \cite{Dresden} and refs. therein),
implying that the sum $\lambda _{p}+\lambda _{s}$ varies between 1.6
and 2.1. W\"{a}lte \textit{et al.} \cite {Dresden} estimated $\omega
_{p}\approx 143$ K, smaller than, but comparable to the calculation
in Ref.\cite{serega}, $\omega _{s}\approx 25$ K and the mass
renormalization due to paramagnons as $1+\lambda _{s}\approx 1.43.$
Then, using  MMF, $\mu ^{\ast }=0.13,$ and $T_{c}=6.8$ K, as
measured for their samples, they deduced $\lambda _{p}$=1.91.

However, there are several problems with this derivation. First of
all, as shown above, the proper formula is Eq.\ref{MM2}. Using this
formula instead of Eq.\ref{MM0}, and keeping all their other
parameters, we get a much more reasonable number, $\lambda
_{p}$=1.61, not far from the value of 1.51 obtained in Ref.
\cite{serega}. However, the SF model adopted in Ref. \cite {Dresden}
cannot be considered as proven. It is based on the disputable
assumption that the upturn of the specific heat quotient at low
temperature and high magnetic field is due to the paramagnon
contribution to specific heat, but there many other explanations of
this effect. 25 K seems to be unrealistically soft. Also, low
$T_{c}$ and high residual resistance cast doubt on the sample
quality in this study.

Here we adopt a different approach: we adopt the calculated values $\lambda
_{p}$=1.5 and $\omega _{p}=131$ K, \ in the harmonic approximation, and
total mass renormalization $1+\lambda _{p}+\lambda _{s}=2.85,$ so that $%
\lambda _{s}=0.35.$ The results of the numerical solution of the
Eliashberg equations with the $\alpha ^{2}F(\omega )$ function
calculated by Ignatov \textit{et al.}\cite{serega} and $\mu ^{\ast
}=0.12$ are shown in Fig .2, together with the curve calculated from
Eq.\ref{Tc}. This way, we find $\omega _{s}\sim 50 $ K, which, we
believe, is a more realistic number than 25 K. The corresponding
total isotope effect coefficient is 0.75.

\begin{figure}[tbp]
\includegraphics[width=2.8in ]{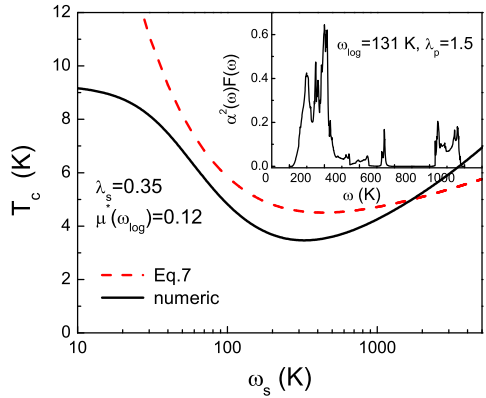}
\caption{$T_c$ for the electron-phonon spectral function calculated
in \protect\cite{serega} for MgCNi$_{3}$ (inset). (color on line)}
\end{figure}

This may sound in agreement with the recent experiment by Klimczuk and Cava %
\cite{Cava}, who have measured the isotope effect to be 0.54 on
carbon only. If the total isotope effect is 0.75, this suggests a
seemingly reasonable Ni isotope effect of 0.21, suggesting that Ni
phonons couple with the electrons twice weaker than C ones.
Unfortunately, the first-principles calculations suggest that the Ni
modes couple with electrons at least an order of magnitude stronger
than the C modes (there is hardly any C character present at the
Fermi level). In the moment, the only way to reconcile this with the
measurements of Ref. \cite{Cava} is to assume that the observed
isotope effect is not a result of the frequency shift of the C
modes, but of some subtle changes in the crystal structure induced
by the isotope substitution. Such a possibility is suggested by an
earlier study \cite{Cavaold}, where it was found that (i) $T_{c}$
depends on the lattice parameter at a rate of $ \approx 310$ K/\AA ,
which translates an error of $\pm 0.0015$ \AA\ in the lattice
parameter\cite{Cava} into an error of $\pm 0.46$ K in $T_{c},$
larger than the isotope shift of 0.3 K, and (ii) that two samples
with the same lattice parameter and the same neutron-measured C
content have $T_{c}$ differing by 0.71 K. A possible explanation is
that, given the proximity of MgCNi$_{3}$ to a ferromagnetic
instability, crystallographic defects may induce local magnetic
moments which, in turn, work as pair-breakers. The concentration of
such defects, even for the same net C content, may depend on the
sample preparation and, possibly, on isotope substitution.

Therefore further studies of the isotope effect both on C and on Ni
are necessary, in particular combined with accurate measurements of
the isotope shift of the phonon modes.

We acknowledge support from the NWO-RFBR grant 047.016.005 and from
the NSF DMR grants 0342290 and 023188.

\end{document}